\documentclass[a4paper,12pt]{article} \usepackage[english]{babel}
\usepackage{fullpage} \usepackage{epsfig} \usepackage{amsfonts}
\usepackage{amssymb} \usepackage{amsmath} \usepackage{theorem}
\usepackage{eufrak}

\numberwithin{equation}{section}

\newtheorem{theorem}{Theorem}[section]
\newtheorem{corollary}[theorem]{Corollary}

\newtheorem{proposition}[theorem]{Proposition} {
\theorembodyfont{\normalfont} \newtheorem{remark}[theorem]{Remark} } {
\theorembodyfont{\normalfont}  }
\makeatletter \title{A Cantor set of tori with monodromy \\ near a focus-focus singularity} \author{Bob Rink\thanks{Mathematics
Institute, Utrecht University,
  PO Box 80.010, 3508 TA Utrecht, The Netherlands, E-mail:
    {\tt rink@math.uu.nl} }} 
\begin{document} \label{monandkam} \hyphenation{boun-da-ry}
\maketitle
\abstract{\noindent We write down an asymptotic expression for action coordinates in an integrable Hamiltonian system with a focus-focus equilibrium. From the singularity in the actions we deduce that the Arnol'd determinant grows infinitely large near the pinched torus. Moreover, we prove that it is possible to globally parametrise the Liouville tori by their frequencies. If one perturbs this integrable system, then the KAM tori form a Whitney smooth family: they can be smoothly interpolated by a torus bundle that is diffeomorphic to the bundle of Liouville tori of the unperturbed integrable system. As is well-known, this bundle of Liouville tori is not trivial. Our result implies that the KAM tori have monodromy. In semi-classical quantum mechanics, quantisation rules select sequences of KAM tori that correspond to quantum levels. Hence a global labeling of quantum levels by two quantum numbers is not possible.}

\section{Introduction}
In this paper we study singular Lagrangean foliations of focus-focus type in two degree of freedom integrable Hamiltonian systems. Such foliations consist of a nontrivial bundle of two-dimensional regular Liouville tori and one singular surface, a pinched torus, see also \cite{Cushman}, \cite{san} or later in this paper. This type of foliation has been found in various two degree of freedom Hamiltonian systems, see for instance \cite{Cushman}, \cite{sadovskii}, \cite{Rink3} and \cite{Sad&zil}. The most famous example is perhaps the spherical pendulum, see \cite{Duistermaat}. \\
\indent Some authors have studied what happens in a perturbation of a Hamiltonian system with such a foliation. Horozov \cite{Horozov} was the first to show that for the spherical pendulum the so-called Arnol'd determinant is nonzero at every regular value of the energy-momentum map. This nondegeneracy condition is traditionally called the Kolmogorov condition and it makes the KAM theorem work. In \cite{Tienzung2} it was proved by Tien Zung that the Kolmogorov condition is satisfied in a full neighbourhood of any pinched torus of focus-focus type. In this paper the results of \cite{Tienzung2} will be made more specific. Based on a computation of Vu Ngoc \cite{san}, we shall explicitly describe the limiting behaviour of the frequency map and the Arnol'd determinant near a pinched torus. The latter grows infinitely large. The Kolmogorov condition requires that the Liouville tori of an integrable system can locally be parametrized by their frequencies. We will see that in the vicinity of a pinched torus, this can even be done more or less globally. \\
\indent What happens to the singular foliation if one perturbs the integrable system? The pinched torus, being a high dimensional homoclinic connection, most likely breaks up into a homoclinic tangle with complinated geometry. A Cantor set of Liouville tori survives as KAM tori. \\
\indent It is well-known, see \cite{Cushman}, \cite{Duistermaat}, \cite{matveev}, \cite{Tienzung}, that the Liouville tori near a pinched torus do not form a trivial torus bundle, but have monodromy. In \cite{sadovskii} the question was posed whether this global geometry remains present in the KAM tori of the perturbed system. In this paper we will show that this is the case. It turns out that the KAM tori in the perturbed system form a Whitney smooth family that is diffeomorphic to the bundle of Liouville tori in the integrable system. This means that the KAM tori have monodromy. The geometry of KAM tori in nearly integrable Hamiltonian systems is also discussed in \cite{BCF}.  The approach in that paper is completely different: the authors use a partition of unity for glueing together local Whitney smooth families of KAM tori.\\
\indent The fact that KAM tori can have monodromy is particularly interesting for semi-classical quantum mechanics. Quantum monodromy in integrable systems has been analysed using the quantum energy-momentum map, see \cite{C&D}, \cite{sadovskii}, \cite{Sad&zil} and \cite{san2}. Semi-classical quantisation theory selects regular sequences of Liouville tori in the classical integrable system which correspond to quantum levels in the quantum system. In the semi-classical limit, with Planck's constant going to zero, these quantum levels form locally a regular lattice of which the points can be labeled by quantised actions. But if monodromy is present in the Liouville tori, then any global labeling of the quantum levels by two quantum numbers is impossible, since there is a shift in the global lattice structure, see \cite{C&D} and \cite{san2}.  \\
\indent  The only problem is that most two degree of freedom Hamiltonian systems are not integrable, even though it might be possible to approximate them by an integrable system. Our result explains that monodromy is also present in nonintegrable systems that are perturbations of an integrable system with a focus-focus singularity. Semi-classical quantisation theory states that the quantum levels are in this case described by sequences of KAM tori, see \cite{Lazutkin}. Monodromy in these KAM tori will again constitute an obvious obstruction to the global labeling of the quantum levels by two quantum numbers. The problems with the global labeling of quantum levels and the phenomenon of redistribution of quantum states have been studied extensively and have also been found experimentally, see for instance \cite{sadovskii}, \cite{overzichtzhil} and \cite{Sad&zil}.
\\
\indent In \cite{Rink3} it was shown that focus-focus equilibria, pinched tori and monodromy can also occur in the Birkhoff normal form of the famous Fermi-Pasta-Ulam (FPU) lattice. Remarkably, their influence could also be observed in numerical integrations of the original lattice equations, even at rather high energy and in lattices of high dimension, where one would usually question the validity of a normal form approximation. This indicates that maybe an exceptional amount of KAM tori will persist in perturbations of integrable                                                                                                                                                                                                                                                                           Hamiltonian systems with focus-focus singularities. A detailed study is necessary to prove this and this paper can be considered as a starting point for such an analysis. \\

\section{Mathematical background}
Let us recall the Hamiltonian monodromy theorem. Let $M$ be a four dimensional real analytic symplectic manifold with symplectic form $\sigma$. Suppose that we have two real analytic Poisson commuting Hamiltonians $H_1, H_2 : M \to \mathbb{R}$, so $\{H_1, H_2\} = 0$. The map $H = (H_1, H_2):M \to \mathbb{R}^2$ is called the energy-momentum map. We want study how the level sets $H^{-1}(h)$ of the energy-momentum map foliate the symplectic manifold $M$. At regular points, this foliation is Lagrangean, because $H_1$ and $H_2$ commute. A point $m \in M$ is called a focus-focus equilibrium if $X_{H_1}(m) = X_{H_2}(m) = 0$ and there are canonical coordinates $(q,p)$ (that is $\sigma = \sum_{i=1}^2 dq_i\wedge dp_i$) near $m$ such that $(q,p)(m)=0$ and
$$H_1 = a (q_1 p_2 - q_2 p_1) + b (q_1 p_1 + q_2 p_2) + \ \mbox{higher order terms}$$
$$H_2 = c (q_1 p_2 - q_2 p_1) + d (q_1 p_1 + q_2 p_2) + \ \mbox{higher order terms}$$
where $a d - b c \neq 0$.  
\noindent Let us moreover assume that $H$ has the following properties:
\begin{list}{}{\leftmargin.7cm \itemsep.15cm \rightmargin.7cm }
\item[1.] There is an open neighbourhood $U \subseteq \mathbb{R}^2$ of $0$ such that $0$ is the only critical value of $H$ in $U$. 
\item[2.] For every $u \in U\backslash \{0\}$, the fiber $H^{-1}(u)$ is connected and compact.
\item[3.] The singular fiber $H^{-1}(0)$ is connected and compact and $m$ is its only singular point. 
\end{list}
The foliation of $H^{-1}(U)$ in level sets of $H$ is a singular Lagrangean foliation with one singular point.
The Arnol'd-Liouville theorem says that the regular fibers $H^{-1}(u)\ (u \in U\backslash \{0\})$ form a smooth bundle of two-dimensional tori. The Hamiltonian monodromy theorem states that this bundle is not trivial. In fact, using a suitable basis for the fundamental group of the torus $H^{-1}(\bar u)$  $(\bar u \in U \backslash\{0\})$ and identifying this torus with the lattice $\mathbb{R}^2 / \mathbb{Z}^2$, the monodromy map of the bundle is given by the matrix 
$\left(\begin{array}{cc} 1 & -1\\0 & 1 \end{array}\right)$. The Hamiltonian monodromy theorem was proved by Matveev \cite{matveev} and Tien Zung \cite{Tienzung}. The monodromy of the bundle is an obvious obstruction to the existence of global action angle coordinates on $H^{-1}(U\backslash \{0\})$, see \cite{Duistermaat}.\\
\indent The singular fiber $H^{-1}(0)$ is a pinched torus: an immersed sphere with one point of transversal self-intersection. Its set of nonsingular points, $H^{-1}(0) \backslash \{m\}$ is diffeomorphic to a cylinder.  \\
\indent Let us write 
$$ K_1 = q_1 p_2 - q_2 p_1 \ \ , \ \  K_2 = q_1 p_1 + q_2 p_2 \ \ \mbox{and} \ K=(K_1, K_2)\ . $$
\noindent Near $m$, the following linearisation result holds and is due to Eliasson \cite{Eliasson}. There exist real analytic canonical coordinates $x=(q, p): W \to T^*\mathbb{R}^2$ in a neighbourhood $W$ of $m$ such that $ x(m) = 0$ and $H = \lambda \circ K \circ x$ for some real analytic local diffeomorphism $\lambda$ of $\mathbb{R}^2$. This means that in $W$, $K \circ x$ and $H$ define the same level sets. We also say that $K \circ x$ is a momentum map for the foliation given by the energy-momentum map $H$. But this implies that $K \circ x$ has a unique extension to a function on $H^{-1}(W)$ that is constant on the level sets of $H$. By the local submersion theorem, this extension is analytic too. We conclude that $K \circ x$ can be extended to a global real analytic momentum map for the Lagrangean foliation near the pinched torus. In other words, since we will only be interested in a neighbourhood of the pinched torus, we can and will assume that $H|_W = K \circ x$.\\ 
\indent Let $F_0: M \to \mathbb{R}$ be an arbitrary real analytic Hamiltonian function which Poisson commutes with $H_1$ and $H_2$, that is $F_0$ is a function of $H_1$ and $H_2$ only. We also write $F_0$ for $F_0 \circ H$. Clearly, the Hamiltonian vector field $X_{F_0}$ is integrable: its flow leaves the level sets of $H$ invariant. The motion in the Liouville tori is simply periodic or quasi-periodic. One may wonder whether this quasi-periodic behaviour persists when we perturb the Hamiltonian function $F_0$. In order to apply the KAM theorem, one must show that the Liouville tori of the integrable system defined by $F_0$ can locally be parametrised by their frequencies. We shall show that this is possible under the assumption that $m$ is a linearly unstable equilibrium point of $X_{F_0}$. \\
\indent Moreover, we want to study the geometry in the KAM tori by giving a smooth torus bundle that interpolates them. Theorems providing interpolation results for KAM tori usually only work for perturbations of integrable systems admitting global action-angle coordinates. This type of theorem will be used at an intermediate stage. This paper then provides an example of a nontrivial Whitney smooth bundle of KAM tori.

\section{Global action-angle coordinates}\label{actie}
\begin{figure}[h] 
\centering \includegraphics[width=10cm, height=4cm]{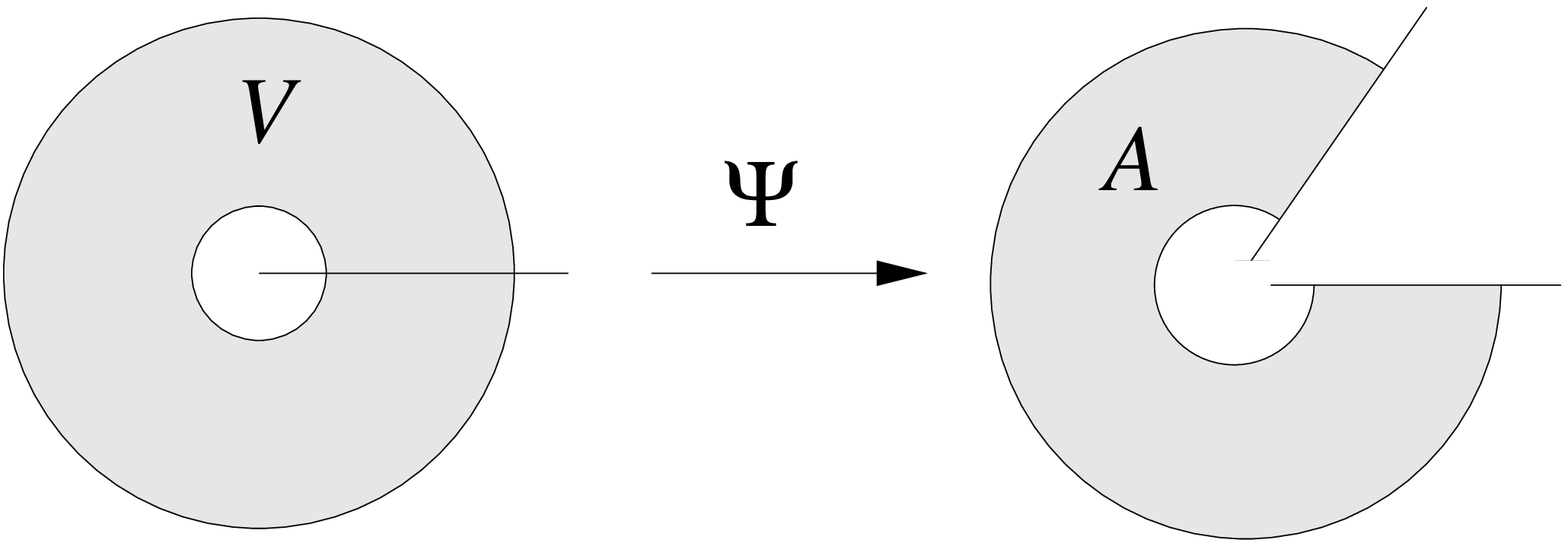} \renewcommand{\figurename}
{\rm \bf \footnotesize Figure}
\caption{\footnotesize The map $\Psi: V_* \to A$.}
\label{psiplaatje}
\end{figure}

Vu Ngoc in \cite{san} derives an expression for action coordinates near a pinched torus from a local analysis near the focus-focus singularity in Eliasson's canonical coordinates $(q, p)$. We recall this result here. \\ 
\indent Let us first define the function $\arg:\mathbb{R}^2 \backslash \{0\} \to \mathbb{R}$ by $\arg:(r\cos\phi, r\sin\phi) \mapsto \phi$. Note that $\arg$ is of course a multivalued function. Over a positively oriented circle around the origin its value increases $2\pi$. But after choosing a fixed branch, $\arg$ becomes locally a uniquely defined analytic function. We have the following
\begin{proposition}
There exists a real analytic function $s = s(H)$ defined on an open neighbourhood $\tilde U \subset \mathbb{R}^2$ of $\{H=0\}$ such that
$$a_1(H) = H_1 \ , \ a_2 (H) =  \frac{1}{2\pi} \left( H_2 \ln |H| + H_1 \arg H \right) + s(H) =: \psi(H)$$
define a set of local action coordinates near every point in $\tilde U\backslash \{0\}$. 
\end{proposition}
\noindent The proof can be found in \cite{san}. Obviously, $a_1(H) = H_1$ ($=K_1$ in Eliasson's coordinates) defines a Hamiltonian vector field in Eliasson's coordinates which has a $2\pi$-periodic flow. Therefore it is an action. The other action is obtained as the Arnol'd integral 
$$a_2(H) = \frac{1}{2\pi} \int_{\gamma_h} \alpha \ ,$$
where $\alpha$ is a one-form such that $d\alpha = \sigma$, see \cite{Duistermaat}. Such a one-form exists locally near every Liouville torus since the foliation in tori is Lagrangean. $\gamma_h$ is a closed curve on the Liouville torus $H^{-1}(h)$ which is chosen so that an integral curve of $X_{a_1}=X_{H_1}$ and  $\gamma_h$ together form a basis of the fundamental group of the torus $H^{-1}(h)$. Obviously, this integral depends analytically on $H$, hence the function $s$ is analytic. In \cite{san} it was shown  that the Taylor expansion of $s-s(0)$ classifies the singular Lagrangean foliation in an open neighbourhood of the pinched torus, up to a symplectomorphism.\\
\indent Let us examine the coordinate transformation $H \mapsto a$ in more detail. We shall for convenience write
\begin{equation}\nonumber
\Psi: (H_1, H_2) \mapsto (a_1, a_2) = (H_1, \psi(H_1, H_2))  
\end{equation}
If we choose a branch of $\arg$, then $\Psi$ is a single valued real analytic map on the domain $\mathbb{R}^2_* : = \mathbb{R}^2 \backslash \mathbb{R}_{\geq 0}(1,0)$ intersected with $\tilde U$. 
The Jacobi determinant of $\Psi$ is  
$$\det D\Psi(H) = \frac{\partial \psi(H)}{\partial H_2} = \frac{1}{2\pi} \left( \ln|H| + 1 \right) + \frac{\partial s(H)}{\partial H_2}$$
which is obviously negative and hence nonzero in a small enough open neighbourhood of $\{H=0\}$. Let us choose a little annulus $V := \{ \rho_1 < |H| < \rho_2\}$ in this neighbourhood. It is easy to verify that $\Psi: V_* := V \backslash \mathbb{R}_{\geq 0}(1,0) \to \mathbb{R}^2$ is injective, because the map $H_2 \mapsto \psi(a_1, H_2)$ has strictly negative derivative and makes a negative jump at $H_2 =0$ if $a_1 >0$. Therefore, $\Psi$ is a real analytic diffeomorphism between $V_*$ and $A:=\Psi(V_*)$. $\Psi$ `opens' $V_*$, that is at different branches of arg, the boundary half line $\mathbb{R}_{\geq 0}(1,0)$ is mapped by $\Psi$ to different half lines, see Figure \ref{psiplaatje}.
 \\
\noindent Let us write $M_V:= H^{-1}(V) \subset M$ for the nontrivial bundle of Liouville tori over $V$. 
\begin{proposition} There exist global action-angle coordinates on the subbundle $H^{-1}(V_*)$  $\subset M_V$. This means that there exist an open set $A \subset \mathbb{R}^2$ and a $C^{\infty}$ diffeomorphism $\Phi_0: H^{-1}(V_*) \to A \times T^2$ with the properties that $\Phi^*_0(da \wedge d\phi) = \sigma$ and $p \circ \Phi_0 = \Psi \circ H$. Here $p:A \times T^2 \to A$ denotes the projection on the first coordinate.
\end{proposition}
{\bf Proof.} Note that $V_*$ is contractible on itself to a point. Therefore the bundle $H^{-1}(V_*)$ over $V_*$ is topologically trivial, see \cite{Steenrod} pp. 53. This implies that there is a homotopy between the identity map on $H^{-1}(V_*)$ and a map that sends $H^{-1}(V_*)$ to a single Lagrangean fiber. Hence $\sigma|_{H^{-1}(V_*)}$ is exact by the homotopy principle. Finally, $\Psi: V_* \to A$ is a diffeomorphism. According to Theorem 2.2 in \cite{Duistermaat} these facts are sufficient for the existence of $C^{\infty}$ global action-angle coordinates on $H^{-1}(V_*)$. $\hfill \square$


\section{Frequencies}\label{freq}
\begin{figure}[h] 
\centering \includegraphics[width=10cm, height=4cm]{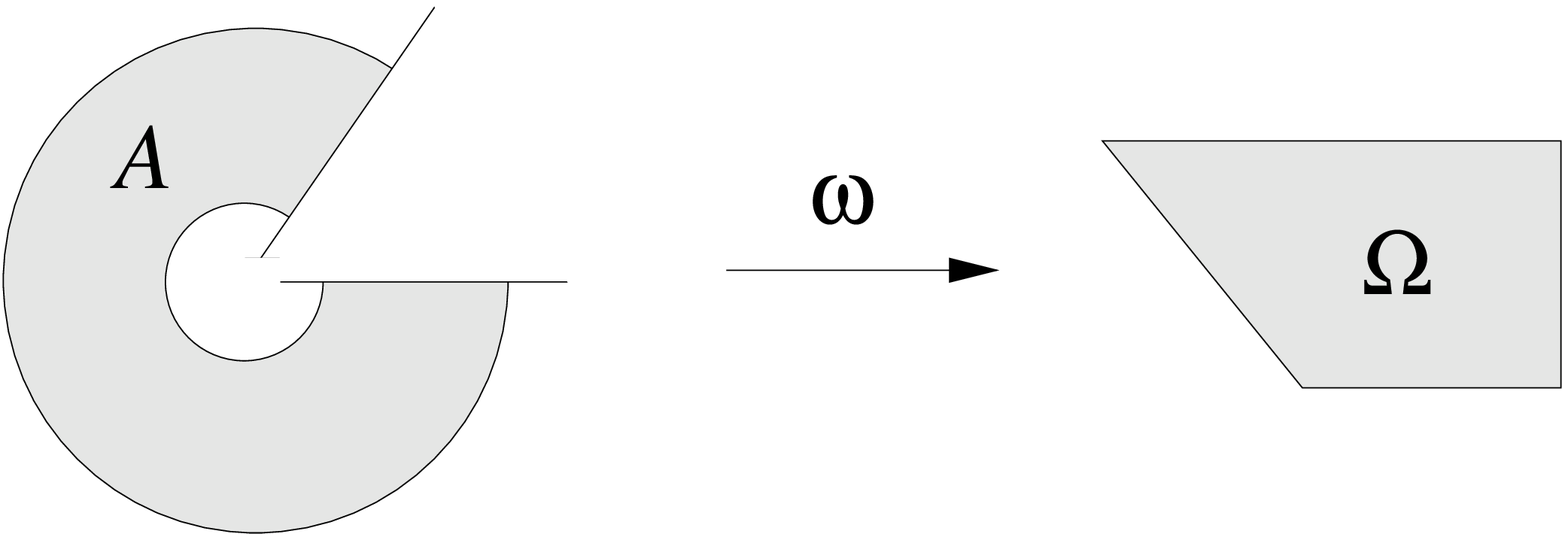} \renewcommand{\figurename}
{\rm \bf \footnotesize Figure}
\caption{\footnotesize The map $\omega: A \to \Omega$.}
\label{blaplaatje}
\end{figure}

\noindent Under the assumption that the focus-focus equilibrium point $m$ is linearly unstable for the vector field $X_{F_0}$, we will show that the frequency map $\omega := \frac{\partial F_0}{\partial a}: A \to \Omega:=\omega(A)$ is a real analytic diffeomorphism. Knowing that $\Psi : V_* \to A$ is a diffeomorphism, we only need to show that $\omega \circ \Psi : V_* \to \Omega$ is a diffeomorphism. We explicitly calculate $\omega \circ \Psi$ as follows. First of all, note that $\omega(a) = \frac{\partial F_0(a)}{\partial a} = \left. \frac{\partial F_0(H)}{\partial H} \right|_{H=H(a)} \frac{\partial H(a)}{\partial a}$ where 
$$\frac{\partial H(a)}{\partial a} = \left(\left.\frac{\partial \Psi(H)}{\partial H}\right|_{H = H(a)}\right)^{-1} = \left( \begin{array}{cc} 1 & 0 \\ \partial\psi(H(a)) / \partial H_1 & \partial\psi(H(a)) / \partial H_2 \end{array}\right)^{-1}\ .$$
Using the fact that 
$$\frac{\partial \psi(H)}{\partial H_1} = \frac{1}{2\pi} \arg H + \frac{\partial s(H)}{\partial H_1}\ \ \ \mbox{and} \ \ \ \frac{\partial \psi(H)}{\partial H_2} = \frac{1}{2\pi} \left( \ln|H| + 1 \right) + \frac{\partial s(H)}{\partial H_2}\ \ ,$$
we arrive at the following expression for $\omega \circ \Psi: H \mapsto \omega(H)$:
\begin{align}\label{omegavanH} \nonumber
\omega_1(H) =& \frac{\partial F_0(H)}{\partial H_1} - \frac{\partial F_0(H)}{\partial H_2} \ \frac{\frac{1}{2\pi} \arg H + \frac{\partial s(H)}{\partial H_1}}{\frac{1}{2\pi} \left( \ln|H| + 1 \right) + \frac{\partial s(H)}{\partial H_2}} \\  \\ \nonumber \omega_2(H) =& \frac{\partial F_0(H)}{\partial H_2} \ \frac{1}{\frac{1}{2\pi} \left( \ln|H| + 1 \right) + \frac{\partial s(H)}{\partial H_2}}
\end{align}
Note that $\lim_{H \to 0} \omega(H) = (\partial F_0(0)/\partial H_1, 0)$. Recall that this limit is taken over $H \in \mathbb{R}^2_*$ and that we have chosen a fixed branch of $\arg$.
The following proposition describes the limiting behaviour of the derivative matrix $\frac{\partial \omega(H)}{\partial H}$ near $H = 0$.
\begin{proposition}
\begin{equation}\label{limiet}
\lim_{H\to 0} \left( \begin{array}{cc} \ln |H| & 0 \\ 0 & \frac{\ln^2|H|}{2\pi} \end{array} \right) \frac{\partial \omega(H)}{\partial H} \left( \begin{array}{cc} H_2 & -H_1 \\ -H_1 & -H_2 \end{array} \right) = \frac{\partial F_0(0)}{\partial H_2} \ \rm{Id}
\end{equation}
\end{proposition}
This follows from a straightforward analysis based on (\ref{omegavanH}). We are now in position to show that $\omega \circ \Psi: V_* \to \Omega$ is a diffeomorphism if $V$ is chosen close enough to the origin $H=0$.
\begin{corollary} $DX_{F_0}(m)$ has an eigenvalue off the imaginary axis if and only if $\frac{\partial F_0(0)}{\partial H_2} \neq 0$. In this case the Arnol'd determinant $\det \left( \frac{\partial \omega(H)}{\partial H} \right)$ goes to infinity as $|H|$ goes to zero. If the annulus $V$ is chosen small enough, then the map $\omega \circ \Psi:V_* \to \Omega$ is a real analytic diffeomorphism. Hence, $\omega: A \to \Omega$ is a real analytic diffeomorphism.
\end{corollary}
{\bf Proof} \ The first statement is trivial since $$DX_{F_0}(m) = \frac{\partial F_0(0)}{\partial H_1}  DX_{H_1}(m) + \frac{\partial F_0 (0)}{\partial H_2} DX_{H_2}(m) \ , $$ and $DX_{H_1}(m)$ and $DX_{H_2}(m)$ commute and respectively have purely imaginary and purely real eigenvalues. The second statement follows by taking the the determinant of (\ref{limiet}) which yields that $$\frac{1}{2\pi}|H|^2\ln^3|H| \det (\frac{\partial \omega(H)}{\partial H}) \to -(\frac{\partial F_0(0)}{H_2})^2 \neq 0$$ and hence $\det (\frac{\partial \omega(H)}{\partial H}) \to \infty$ as $H \to 0$. According to Proposition \ref{limiet} we can now choose the annulus $V$ in such a way that for every $H \in V_*$,
\begin{equation}\label{limiet2}\nonumber 
\left( \begin{array}{cc} \ln |H| & 0 \\ 0 & \frac{\ln^2|H|}{2\pi} \end{array} \right) \frac{\partial \omega(H)}{\partial H} \left( \begin{array}{cc} H_2 & -H_1 \\ -H_1 & -H_2 \end{array} \right) = \frac{\partial F_0(0)}{\partial H_2} \ \left(  \ {\rm Id} + M(H) \ \right) 
\end{equation} 
\noindent \hskip-.135cm for some matrix $M(H)$ of which the elements each have norm less than $\frac{1}{10}$. This clearly implies that $\det (\frac{\partial \omega(H)}{\partial H}) \neq 0$ for $H \in V_*$. It is easy to show that this implies that $\omega \circ \Psi: V_* \to \Omega$ is injective. Pick $H^{(1)}$ and $H^{(2)}$ in $V_*$. We connnect $H^{(1)}$ and $H^{(2)}$ by a curve $\gamma$ consisting of a circle segment from $H^{(1)}$ to $\frac{|H^{(1)}|}{|H^{(2)}|}H^{(2)}$ and a line segment from $\frac{|H^{(1)}|}{|H^{(2)}|}H^{(2)}$ to $H^{(2)}$. A straightforward but rather long computation shows that $\omega(H^{(2)}) - \omega(H^{(1)}) = \int_{\gamma} \frac{\partial \omega}{\partial H}\cdot d s \neq 0$, expressing that $\omega \circ \Psi$ is injective. This proves that $\omega \circ \Psi$ and $\omega$ are real analytic diffeomorphisms.  $\hfill \square$
\\
\\
We conclude that if $DX_{F_0}(m)$ has an eigenvalue with nonzero real part, then it is possible to choose the annulus $V$ close enough to the origin $H=0$ such that both the action map $\Psi: V_* \to A$ and the frequency map $\omega: A \to \Omega$ are diffeomorphisms. \\
\indent $V_*, A$ and $\Omega$ are open, contractible, bounded sets with a piecewise smooth boundary, see Figures \ref{psiplaatje} and \ref{blaplaatje}. 

\section{Monodromy in the KAM tori}
We shall argue that if one perturbs the completely integrable Hamiltonian $F_0$ on $M$ a bit, the monodromy of the Liouville tori in $M_V$ is still present in the surviving KAM tori. It turns out that the KAM tori form a Whitney smooth torus bundle. They can be interpolated by a smooth torus bundle that is diffeomorphic to $M_V$. This type of interpolation theorem is well-known for perturbations of an integrable Hamiltonian system for which 
\begin{list}{}{\leftmargin.7cm \itemsep.15cm \rightmargin.7cm }
\item[1.] global action-angle coordinates exist.
\item[2.] the frequency map is a global diffeomorphism. 
\end{list} 
\noindent See \cite{Broer} or \cite{Poschel}. Obviously, $M_V$ does not meet these requirements, since it is not a trivial bundle. According to the previous paragraphs, the bundle $H^{-1}(V_*) \subset M_V$ does satisfy 1. and 2. We now apply the standard KAM theorem on this subbundle and see that this suffices to get an interpolation result for all the KAM tori in $M_V$.   \\
\indent A well-known KAM interpolation theorem is for instance given in \cite{Poschel} by P\"oschel. The setting of P\"oschel's theorem is the following:\\
\indent Let $A \subset \mathbb{R}^2$ be an open subset. Consider the symplectic manifold $A \times T^2$ with symplectic form $da \wedge d\phi$ and real analytic Hamiltonian function $\widetilde{F}_0(a)$. Assume that $\omega = \frac{\partial \widetilde{F}_0}{\partial a}: A \to \mathbb{R}^2$ is a global diffeomorphism on its image, which is the set of frequencies $\Omega$. Obviously, under the diffeomorphism  $\Phi_1= \omega^{-1} \times \mbox{Id}: \Omega \times T^2 \to A \times T^2$, the
Hamiltonian vector field $X_{\widetilde{F}_0} = \frac{\partial \widetilde{F}_0}{\partial a_1 }\frac{\partial }{\partial \phi_1 } + \frac{\partial \widetilde{F}_0}{\partial a_2 }\frac{\partial }{\partial \phi_2}$ on $A \times T^2$ pulls back to the vector field
$$\Phi_1^*X_{\widetilde{F}_0} = \omega_1 \frac{\partial}{\partial \phi_1} + \omega_2 \frac{\partial}{\partial \phi_2}$$ 
on $\Omega \times T^2$. P\"oschel's theorem says the following for perturbations of $X_{F_0}$:
\begin{theorem}[KAM]\label{poschel}
Let $\tau > 1$ be a fixed given number. Then there exists a positive constant $\delta$ such that for every small enough $\gamma$ and every $C^{\infty}$ Hamiltonian function $\widetilde{F}(a, \phi)$ with 
$$||\widetilde{F} - \widetilde{F}_0|| < \delta \gamma^2 $$
the following holds: there exists a $C^{\infty}$ near identity diffeomorphism $\Phi_2: \Omega \times T^2 \to \Omega \times T^2$  which on $\Omega_{\gamma}\times T^2$ conjugates the vector field $\Phi_1^*X_{\widetilde{F}}$ to the vector field $\Phi_1^*X_{\widetilde{F}_0}$, that is 
$$\Phi^*_2 \Phi^*_1 X_{\widetilde{F}}\left| _{\Omega_{\gamma} \times T^2} \right.  = (\Phi_1 \circ \Phi_2)^*X_{\widetilde{F}} \left| _{\Omega_{\gamma} \times T^2} \right. = \Phi_1^* X_{\widetilde{F}_0} $$
Here $\Omega_{\gamma}$ is defined as the set of frequencies $\omega \in \Omega$ that have distance at least $\gamma$ to the boundary of $\Omega$ and satisfy the Diophantine inequalities $$|(\omega, k)| \geq \gamma |k|^{-\tau} \ \ \forall \  k \in \mathbb{Z}^2$$
By construction, $\Phi_2(\omega, \phi) = (\omega, \phi)$ if $\omega$ has distance less than $\gamma/2$ to $\partial \Omega$.
\end{theorem}
\begin{remark}
The norm $|| \cdot ||$ is a combination of a $C^{\infty}$ supremum norm and a $C^{\infty}$ H\"older norm for smooth functions on $A \times T^2$, see \cite{Poschel} pp. 662-663 and 690.
\end{remark}
\begin{remark} P\"oschel also assumes that $\widetilde{F}_0$ has a complex analytic extension to a neighbourhood of $A$ in $\mathbb{C}^2$. We avoid this problem by switching to a smaller $A$ if necessary, which in our case can be arranged by choosing the annulus $V$ appropriately.
\end{remark}
\begin{remark} P\"oschel uses the freedom in the Whitney extension theorem to construct $\Phi_2$ such that $\Phi_2(\omega, \phi) = (\omega, \phi)$ if $\omega$ has distance less than $\gamma/2$ to $\partial \Omega$, see \cite{Poschel} pp. 681-682. This has the effect that $\Phi_2$ becomes a $C^{\infty}$ diffeomorphism from $\Omega \times T^2$ to $\Omega \times T^2$. We will see that the same property has more advantages.
\end{remark}
\begin{remark}
The domain $\Omega$ in section \ref{freq} of this paper is bounded and has a piecewise smooth boundary. Therefore one quickly derives from the definition of $\Omega_{\gamma}$ that the Lebesgue measure of $\Omega\backslash \Omega_{\gamma}$ is of order $\gamma$. This means that there are positive constants $L$ and $\gamma_0$ such that the Lebesgue measure of $\Omega \backslash \Omega_{\gamma}$ is smaller then $L \gamma$ if $\gamma < \gamma_0$.
\end{remark} 
Theorem \ref{poschel} says that the tori with frequencies $\omega$ in the Cantor set $\Omega_{\gamma} \subset \Omega$ survive a small enough Hamiltonian perturbation. The surviving KAM tori form a Whitney smooth family of tori, that is they can be interpolated by a smooth bundle of tori that lie close to the original tori. \\
\\
We can now simply apply P\"oschel's theorem for perturbations of a Hamiltonian system on $A \times T^2$, where $A=\Psi(V_*)$. Then it is only a small step to our main theorem:

\begin{theorem} \label{kamenmonodromie} Let $V$ be an annulus around $\{H=0\}$ such that $\Psi$ and $\omega$ are real analytic diffeomorphisms on $V_*$ and $A=\Psi(V_*)$. According to the results of sections  \ref{actie} and \ref{freq}, such a $V$ exists if $DX_{F_0}(m)$ has an eigenvalue that is not purely imaginary. Then there exists a positive constant $\delta$ such that for every small enough $\gamma$ and every $C^{\infty}$ Hamiltonian function $F: M_V \to \mathbb{R}$ with 
$$||(F - F_0) \circ \Phi_0^{-1} || < \delta \gamma^2 $$
the following holds: there exists a $C^{\infty}$ near identity diffeomorphism $\Phi: M_V \to M_V$ and a Cantor set $V_{\gamma} \subset V$ such that 
$$\Phi^*X_F\left|_{H^{-1}(V_{\gamma})}\right. = X_{F_0} \ .$$
The Lebesgue measure of $V \backslash V_{\gamma}$ is of order $\gamma$.
\end{theorem}
{\bf Proof.} For the given Hamiltonians $F_0, F: M_V \to \mathbb{R}$, define $\widetilde{F}_0 = F_0 \circ \Phi_0^{-1} = F_0 \circ \Psi^{-1}$ and $\widetilde{F} = F \circ \Phi_0^{-1}$ on $A \times T^2$. Note that $\widetilde{F}_0$ is analytic if $F_0$ is analytic, since $\Psi$ is analytic. According to P\"oschel's theorem, there is a positive constant $\delta$ such that if $||\widetilde{F} - \widetilde{F}_0|| < \delta \gamma^2$, then there exists a near identity transformation $\Phi_2: \Omega \times T^2 \to \Omega \times T^2$ such that 
$$\Phi^*_2 \Phi^*_1 X_{\widetilde{F}}\left| _{\Omega_{\gamma} \times T^2} \right. = \Phi_1^* X_{\widetilde{F}_0} $$
But because $\Phi_0$ is symplectic, $X_{F_0} = \Phi_0^*X_{\widetilde{F}_0}$ and $X_{F} = \Phi_0^*X_{\widetilde{F}}$. It follows that
$$\Phi^*_2 \Phi^*_1 (\Phi_0^{-1})^* X_{F}\left| _{\Omega_{\gamma} \times T^2} \right. = \Phi_1^* (\Phi_0^{-1})^*X_{F_0} $$
and hence
$$(\ \Phi_0^{-1} \circ \Phi_1 \circ \Phi_2 \circ \Phi_1^{-1} \circ \Phi_0\ )^* X_{F}\left| _{H^{-1}(V_{\gamma})} \right. = \Phi_0^* (\Phi^{-1}_1)^* \Phi^*_2 \Phi^*_1 (\Phi_0^{-1})^* X_{F}\left| _{H^{-1}(V_{\gamma})} \right. = X_{F_0} $$
where $V_{\gamma}$ is defined as $V_{\gamma} := (\omega \circ \Psi)^{-1}(\Omega_{\gamma})$. Thus, $H^{-1}(V_{\gamma}) = (\Phi_1^{-1}\circ \Phi_0)^{-1}(\Omega_{\gamma} \times T^2)$. The Lebesgue measure of $\Omega\backslash \Omega_{\gamma}$ is of order $\gamma$. Because the Jacobi determinant of $(\omega \circ \Psi)^{-1}$ is bounded on $\Omega$, the Lebesgue measure of $V \backslash V_{\gamma}$ is of order $\gamma$ too.\\
 \indent Let us now define the map $\Phi:M_V \to M_V$ as follows: 
$$\Phi (m) = \left\{ \begin{array}{l} 
 (\ \Phi_0^{-1} \circ \Phi_1 \circ \Phi_2 \circ \Phi_1^{-1} \circ \Phi_0\ )(m) \ , \ \mbox{if} \ m \in H^{-1}(V_*) \\
m \ , \ \mbox{if} \ m \in H^{-1}(\mathbb{R}_{\geq 0}(1,0)) 
\end{array} \right.$$
Because $\Phi_2(\omega, \phi) = (\omega,\phi)$ in an open neighbourhood of $\partial \Omega$, $\Phi$ has the property that in a full neighbourhood of the set $H^{-1}(V \cap \mathbb{R}_{\geq 0}(1,0))$ it is the identity map. Furthermore, $\Phi|_{H^{-1}(V_*)} : H^{-1}(V_*) \to H^{-1}(V_*)$ is a diffeomorphism. Hence, $\Phi$ is a diffeomorphism. As we already argued, it has the required conjugation property on $H^{-1}(V_{\gamma})$.
 $\hfill \square$\\
\\
\noindent The diffeomorphism $\Phi: M_V \to M_V$ in Theorem \ref{kamenmonodromie} maps the Liouville torus $H^{-1}(v)$, $(v \in V_{\gamma})$ of the unperturbed integrable system defined by $F_0$ to a KAM torus of the perturbed system defined by $F$. This means that the KAM tori can be interpolated by a family of tori that is diffeomorphic to $M_V$: the KAM-tori have monodromy.

\section{Discussion}
The results obtained in this paper generalize \cite{Tienzung2} in which it is proved that the Kolmogorov condition is satisfied near a focus-focus singular value. We obtain explicit quantitative estimates on the behaviour of the Arnol'd determinant near a pinched torus and show that it grows to infinity. But there is more. By cutting away a measure zero set of Liouville tori, we have obtained a trivial torus bundle on which global action-angle coordinates exist. The tori in this bundle can globally be parametrized by their frequencies. This enables us to use the standard KAM theorem, which says that in a perturbed system certain tori survive and that these tori are part of a smooth structure. By a simple gluing argument, we show that the KAM tori near a pinched torus form a nontrivial Whitney smooth bundle that is diffeomorphic to the original bundle of Liouville tori. This justifies the statement that the KAM tori in a perturbed singular foliation of focus-focus type have monodromy.

\section{Acknowledgement}
The author thanks Henk Broer, Richard Cushman, Hans Duistermaat, Giovanni Gallavotti and Ferdinand Verhulst for challenging discussions and fruitful remarks.

\bibliography{/user1/home7/Rink/bibliografie/bibliografie}
\bibliographystyle{amsplain}

\end{document}